\documentstyle[preprint,pre,aps,eqsecnum,amssymb,epsfig]{revtex}

\begin{document}

\draft
\tightenlines
\title{Shapes, contact angles, and line tensions of
  droplets on cylinders} 
\author{C. Bauer and S. Dietrich}
\address{Fachbereich Physik, Bergische Universit\"at Wuppertal,
D-42097 Wuppertal, Germany}
%\date{\today}

\maketitle

\begin{abstract}
Using an interface displacement model we calculate the shapes of
nanometer-size liquid droplets on homogeneous cylindrical surfaces. We
determine effective contact angles and line tensions, the latter defined as excess  
free energies per unit length associated with the two contact lines at the
ends of the droplet. The dependences of these quantities on
the cylinder radius and on the volume of the droplets are analyzed.
\end{abstract}


\section{Introduction}

The wetting properties of a fiber in liquid matrices (e.g., dye
mixtures, polymer melts, or molten resins) play an important role in the textile
industry and in the fabrication of high-performance, fiber-reinforced
composite materials. Since contact angles of liquid droplets on solid
substrates provide a valuable characterization of such wetting
properties there are numerous experimental and theoretical studies of
the shape and the spreading of droplets deposited on a cylindrical
substrate (see, e.g.,
Refs.~\cite{carroll1,carroll2,carroll4,brochardetal,quereetal,rooksetal,wagner,yarinetal,leechiao,connoretal,mchaleetal,debruyn,songetal,guli,jenkinsdonald}).
The morphology of liquid drops on a fiber is particularly interesting
insofar as on a planar substrate there is only one, spherical caplike
droplet shape, whereas
on a cylindrical substrate droplets may exhibit two, topologically
different shapes, a
``clamshell''- and a ``barrel''-type one, depending on the droplet
volume, the contact angle, and the cylinder
radius~\cite{carroll1,carroll2,carroll4}. The aforementioned studies
deal with thick fibers and large drops, i.e., the length scales are
$\mu$m and larger. In this range the fluid structures are determined
by macroscopic properties alone, i.e., volume of the liquid, surface
tension $\sigma$ of the liquid vapor interface, Young's contact angle
$\theta_{\infty}$, and radius $R$ of the cylinder.

However, with the discovery of nanotubes the interest in such fluid
structures has shifted to much smaller scales. There are several
applications for which these small solid-fluid structures are very
important. (i) For fabricating valuable composite materials involving
nano\-tubes their wetting by the liquid host matrix is necessary to
couple the inherent strength of the nanotubes to the matrix,
reinforcing materials or fillers for plastics and
ceramics~\cite{dujardinetal}. (ii) Nanotubes can be used as supports
for heterogeneous catalysis or as templates for creating small wires
or tubular structures by coating them with metals or metal oxides in
the liquid state~\cite{ajayanetal} or by attaching inorganic and
organic moieties to the nanotube surfaces~\cite{ebbesenetal}. (iii)
In order to use nanotubes as ``nanostraws'' potential candidates for
exploiting such capillarity must be screened by first seeing if the
liquid wets the \emph{outside} of nanotubes~\cite{dujardinetal2}. The
performance of the nanotubes as catalysts, adsorbants, and deodorants
can vary depending on whether they are composed of carbon, boron
nitride, or oxides (SiO$_2$, Al$_2$O$_3$, V$_2$O$_5$, MoO$_3$,
TiO$_2$)~\cite{kasugaetal}. This variety demonstrates, that the
substrate potential of these tubes can be regarded as a tunable
parameter. (iv) By using nanotubes as nanotweezers~\cite{kimlieber} it
might be possible to grab and manipulate small liquid drops. For this
application the substrate must be nonwettable.

These small scales are comparable with the range of the substrate
potential of the cylinders and of the molecular forces between the
fluid particles adsorbing on them. Thus the droplets form under the
action of the so-called effective interface potential $\omega$ which
accounts for the net effect of the competition between the forces
among the fluid particles and the substrate
potential~\cite{sdreview}. Accordingly the calculation of the
corresponding deformed droplet shapes requires a more detailed
theoretical description which takes the effective interface potential
into account. To our knowledge there is only one, recent publication
in which this effect of $\omega$ on the droplet shape on fibers has
been analyzed~\cite{neimark}. It is the purpose of our study here to
refine and to extend this analysis in various directions. If the
radius $R$ of the fiber reduces to a few nm, as it is the case for
nanotubes, the effective interface potential itself will depend on $R$
and thus deviate from that of the corresponding semi-infinite planar
substrate used in Ref.~\cite{neimark}. Accordingly we present a
systematic analysis of the dependence of the shape of the droplets and
their suitably defined contact angles on both $R$ and the droplet
volume. This enables us to describe systematically the crossover in
shape and contact angle between those of droplets on a cylinder and
on the limiting case $R\to\infty$ of a planar substrate. We remark on
how the structure of the effective interface potential, depending on
whether it leads to first-order or continuous wetting transitions,
influences the morphology of the droplets. We confine our analysis to barrel-type 
droplets and estimate their metastability against roll-up to the
clamshell configuration. Finally we study two types of line
tensions. The first one concerns the line tension of three-phase
contact between liquid, vapor, and substrate emerging at the ends of
macroscopicly large drops on fibers which reduces to the familiar line
tension of the straight three-phase contact line on a planar
substrate. The second excess free energy concerns the effective line
tension associated with the circular shape of the three-phase contact
line on a planar substrate as function of the droplet volume. These
results are relevant for understanding how to extract line tensions
from contact angle measurements.

We are encouraged to present our refined analyses by recent
experimental advances to determine droplet shapes such as microscopic
interferometry~\cite{wangbetelulaw}, ellipsometric
microscopy~\cite{neumaieretal}, scanning 
polarization force microscopy~\cite{rieutordsalmeron}, and
tapping-mode scanning force 
microscopy~\cite{herminghausetal}. These techniques allow one to resolve drop
profiles on the submicrometer
scale~\cite{neumaieretal,rieutordsalmeron} down to the nanometer 
scale~\cite{herminghausetal}, both vertically and laterally. In view
of the numerous important applications mentioned above it would be
rather rewarding to extend the application of these techniques to
nonplanar substrate geometries in order to resolve experimentally the
shape of droplets on fibers and tubes as presented in the following
sections. 

\section{Theory}
\label{s:theory}

\subsection{Free energy functional}
\label{ss:func}

In cylindrical coordinates 
the droplet surface is described by a function $h(z)$ or $l(z)$ of the
coordinate $z$ along the symmetry axis of the cylinder
(Fig.~\ref{f:system}(a)). We define $h(z)$ and $l(z)$ 
such that $h(z)$ is the local separation between the liquid-vapor
interface and the symmetry axis of the cylinder and $l(z) = h(z)-R$ is
the local separation between the cylinder surface and the liquid-vapor
interface, i.e., the liquid layer thickness. The droplet is also
symmetric with respect to a reflection at the plane $z=0$. For large
values of $|z|$, i.e., at large distances
from the droplet center at $z=0$, the liquid forms a thin wetting layer of
thickness $l_0 = h_0-R$ around the cylinder. For reasons of simplicity
$h(z)$ is henceforth assumed to be a unique function of $z$, i.e., we
do not consider contact angles $\theta>90^{\circ}$. The shape of the
liquid-vapor interface enclosing the droplet is determined by the
interplay of three physical quantitites: the Laplace pressure~\cite{rowlinsonwidom}
$2\sigma H$ generated by the mean curvature $H$ of the interface with
surface tension $\sigma$, the capillary pressure~\cite{rowlinsonwidom}
induced by the finite droplet volume, and the 
disjoining pressure or, equivalently, the
effective interface potential $\omega_c$ acting on the liquid-vapor
interface~\cite{sdreview}; $\omega_c(l;R)$ is the cost in 
free energy per surface area to maintain a homogeneous wetting layer
of prescribed thickness $l$ covering the cylinder surface and can be
expressed in terms of the underlying forces of the substrate and
between the fluid particles~\cite{bieker}. In the absence of
the effective interface potential, i.e., for large droplets the liquid-vapor
interface is a minimal surface under the constraint of a prescribed
volume, i.e., it exhibits a constant mean curvature. The
influence of the effective interface potential is most pronounced near
the cylinder surface within the range of the substrate potential and
leads to a deviation of the actual profile $h(z)$ from the 
shape which is determined by the aforementioned constant mean curvature condition. 
On the other hand, in the limit of large separation from the cylinder
surface the mean curvature is asymptotically constant because there
the influence of the effective interface potential 
vanishes. 

Independent of the size of the droplet, for later purposes we define the
``reference configuration'' (see Fig.~\ref{f:system}(a)) 
\begin{equation}\label{e:ref}
a_{ref}(z) = a(z)\Theta(z_1-|z|)+h_0\Theta(|z|-z_1)
\end{equation}
where $a(z)$ is that constant-mean-curvature surface which touches the
surface $h(z)$, $h(z=0)=a(z=0)$, and exhibits the
same curvature $H$ at the apex, i.e., the two principal radii of
curvature 
\begin{equation}\label{e:R0}
R_0 = \frac{(1+(h'(0))^2)^{3/2}}{h''(0)} =
\frac{1}{h''(0)}\quad\mbox{with}\quad h'\equiv\frac{dh}{dz} 
\end{equation}
and $h(0) = R+l(0)$ of $h(z)$ and, correspondingly, of $a(z)$ (see
Fig.~\ref{f:system}(a)) at the apex position $z=0$ are
identical. $\Theta$ denotes the 
Heaviside step function; $\pm z_1$ are those values of $z$ where $a(z)$ and
the homogeneous wetting layer $h(z)\equiv h_0$ intersect. In this sense
the values $z=\pm z_1$ define the positions of the two three-phase contact lines
forming the ends of the droplet.
The ``apparent contact angle'' $\theta$ is defined by the intersection
of the barrel-shaped part $a(z)$ of the reference profile and the
homogeneous wetting layer $h(z)\equiv h_0$ (see Fig.~\ref{f:system}(a)):
\begin{equation}\label{e:ca}
\theta = \lim_{z\nearrow z_1}\arctan(|a'(z)|).
\end{equation}
This apparent contact angle
$\theta$ can be expressed~\cite{carroll1,carroll2} in terms of the measurable
quantities apex height $l(0)$ of the droplet, radius $R$ of the
cylinder, radius $R+l_0$ of the wetting film, and radius of curvature
$R_0$ (Eq.~(\ref{e:R0})) of the profile $h(z)$ at the apex:
\begin{equation}\label{e:cacarroll}
\cos\theta = \frac{R+l(0)}{R+l_0}-\frac{(R+l(0))^2-(R+l_0)^2}{2(R+l_0)}
\left(\frac{1}{R+l(0)}+\frac{1}{R_0}\right).
\end{equation}

Within an interface displacement model (see, e.g.,
Ref.~\cite{indekeu}) the equilibrium
interface configuration $\bar{h}(z)$ for a droplet of prescribed
\emph{excess} volume $V_{ex}$ minimizes the free energy functional
\begin{eqnarray}\label{e:func}
F_{ex}[h(z)] & = & F[h(z)]-F[h_0] \\
& = & 2\pi\int\limits_{-\infty}^{\infty}dz\,
\Bigg(\sigma\,\Big(h(z)\,\sqrt{1+h'(z)^2}-h_0\Big)
+ R\,\Big(\omega_c(h(z)-R)-\omega_c(h_0-R)\Big)\Bigg) \nonumber
\end{eqnarray}
under the constraint
\begin{equation}\label{e:constraint}
\pi\int\limits_{-\infty}^{\infty}dz\,\left(h^2(z)-h_0^2\right) =
V_{ex}
\end{equation}
and the boundary conditions $h(|z|\to\infty) = h_0$. We have defined
$F_{ex}[h(z)]$ as an excess free energy with 
respect to the free energy $F[h_0]$ of the homogeneous wetting layer
$h(z)\equiv h_0$ rendering a mathematically well-defined, finite
expression. The first contribution to
$F_{ex}$ is the excess free energy due to the increase of the
liquid-vapor interface as compared with a homogeneous cylindrical
shape. In general $\sigma$ itself depends on the curvature and thus on
$R$ (see, e.g., Sec.~2.2 in Ref.~\cite{bieker} and references
therein); in the following, however, we do not discuss explicitly this
additional parametric dependence on $R$.
The second contribution to $F_{ex}$ is the free energy generated by the
effective interaction between the cylinder surface and the
liquid-vapor interface, reduced by the corresponding free energy for
the homogeneous wetting layer. Since the substrate is considered to be
homogeneous, $\omega_c(l)$ depends only on the radial distance $l =
h-R$ from the substrate surface. The equilibrium separation $h_0 =
l_0+R$ of the homogeneous wetting layer from the cylinder axis
minimizes the free energy
$F(h) = 2\pi L(R\omega_c(h-R)+\sigma\,h)$ where $L\gg z_1$ is the
macroscopic length of the cylinder. The constrained minimum of
Eq.~(\ref{e:func}) is given by the unconstrained minimum of the
surrogate functional
\begin{equation}\label{e:func_constraint}
{\mathcal F}_{ex}[h(z)] =
F_{ex}[h(z)]+\kappa\left(\pi\int\limits_{-\infty}^{\infty}dz\,
  \left(h^2(z)-h_0^2\right)-V_{ex}\right).
\end{equation}
The corresponding optimal profile $\tilde{h}(z,\kappa)$ renders the
equilibrium profile $\bar{h}(z,V_{ex}) = \tilde{h}(z,\kappa(V_{ex}))$
upon expressing the Lagrange multiplier $\kappa$ in terms of $V_{ex}$
by inserting $\tilde{h}(z,\kappa)$ into the left hand side of
Eq.~(\ref{e:constraint}) which yields the implicit relation
$V_{ex}(\kappa)$. In order to avoid a clumsy notation, in the
following we denote $\bar{h}(z,V_{ex})$ by $h(z)$.
The Euler-Lagrange equation
corresponding to Eq.~(\ref{e:func_constraint}) reads
\begin{equation}\label{e:ele}
\sigma\left(\frac{1}{h(z)(1+h'(z)^2)^{1/2}} -
  \frac{h''(z)}{(1+h'(z)^2)^{3/2}}\right) = 2\sigma H(z) = -\kappa -
  \frac{R}{h(z)}\left. \frac{d\omega_c(h-R)}{dh}\right|_{h=h(z)}.
\end{equation}
This equation describes the balance between the Laplace pressure on
the left hand side and the capillary plus disjoining pressure on the
right hand side. One has $\kappa<0$ for any barrel-shaped droplet.

\subsection{Reference profiles}

The reference profile $a(z)$ minimizes a similar surrogate functional:
\begin{equation}\label{e:minsurf}
{\mathcal A}[a(z)] = \pi\int\limits_{-z_1}^{z_1}
dz\,\Big(2\,\sigma\,a(z)\,\sqrt{1+a'(z)^2}+\kappa^*\,a^2(z)\Big) 
+\mbox{const}
\end{equation}
with the constant independent of $a(z)$ and the boundary conditions
$a(\pm z_1) = h_0$.
Equation~(\ref{e:minsurf}) follows from Eq.~(\ref{e:func_constraint}) by
omitting $\omega_c$ and replacing $\kappa$ by $\kappa^*$. The corresponding
Euler-Lagrange equation describes the constant-mean-curvature surface
given by
\begin{equation}\label{e:eleasymp}
\sigma\left(\frac{1}{a(z)(1+a'(z)^2)^{1/2}} -
  \frac{a''(z)}{(1+a'(z)^2)^{3/2}}\right) = 2\sigma H(z) = -\kappa^*.
\end{equation}
According to the definition of $a(z)$ the Lagrange multiplier
$-\kappa^*$ has to be chosen such that this constant mean curvature of
this surface equals the mean curvature at the apex of the actual
surface $h(z) = l(z)+R$:
\begin{equation}\label{e:cmc}
-\frac{\kappa^*}{\sigma} = \frac{1}{R+l(0)}+\frac{1}{R_0}
\end{equation}
$R_0$ (see Eq.~(\ref{e:R0})) and $R+l(0)$ are the principal radii of
curvature at the apex of the actual
surface $h(z)$, determined by the former Lagrange multiplier $\kappa(V_{ex})$.
The solution of Eq.~(\ref{e:eleasymp}) is given implicitly by
\begin{equation}\label{e:asymp}
\int\limits_{h_0}^{a(z)}
dy\left(\left(\frac{\sigma}{\kappa^*}\right)^2\left(\frac{2y}
{{\mathcal C}-y^2}\right)^2-1\right)^{-1/2} = z+z_1, \qquad |z|\leq z_1,  
\end{equation}
which fulfils the boundary condition $a(-z_1) = h_0$; this determines
implicitly $z_1$ in terms of $h_0$, $V_{ex}$, $\sigma$, and
$\omega_c$. The integration constant ${\mathcal C}$ 
is determined by $a'(0)=0$ due to the symmetry of $a(z)$. The integral
in Eq.~(\ref{e:asymp}) can be expressed in terms of elliptic
integrals~\cite{carroll3,yamakikatayama}.

When the drop is macroscopicly large ($V_{ex}=\infty$) it is
appropriate to adopt a slightly different point of 
view. In this case not the center of the droplet but the position of one of the two
three-phase contact lines, which are defined by the intersection of the asymptote
$a_m(z)$ (the constant-mean-curvature surface appertaining to the
\emph{m}acroscopic drop) and $h(z)\equiv h_0$, is fixed at $z=0$ (see
Fig.~\ref{f:system}(b)). The actual interface profile interpolates between,
e.g., $h(z\to-\infty)=h_0$ and $h(z\to\infty) = a_m(z)$. This configuration 
describes a solid cylinder which is in contact with bulk vapor on the left hand
side ($z\to-\infty$) and with bulk liquid on the right hand side
($z\to\infty$). The analysis of the internal structure of a
three-phase contact line on a 
homogeneous, planar substrate is based on a similar configuration
(see Refs.~\cite{getta} and \cite{bauer1} and references therein). The interface
profile diverges in the limit $z\to\infty$: $h(z\to\infty)\to\infty$,
but this divergence is not linear as
in the case of the planar substrate. A macroscopicly large drop
implies $\kappa\to0$. In this limit the volume constraint loses its
meaning. Instead the state of the system is fixed by
different lateral boundary conditions. In this case the
solution of Eq.~(\ref{e:asymp}) is given by
\begin{equation}\label{e:asympmacroCandD}
a_m(z) = a(z;\kappa=0) = C\,\mbox{cosh}\left(\frac{z-D}{C}\right)
\end{equation}
with two integration constants $C$ and $D$. $a_m(z)$ describes a
rotational surface with minimal surface area. From
Eq.~(\ref{e:asympmacroCandD}) one can easily see that the divergence of the  
interface profile for macroscopic drops is \emph{exponential},
$a_m(z\to\infty)=\frac{C}{2}\exp((z-D)/C)$, rather than linear as on a
planar substrate. The reference 
profile appertaining to the macroscopic drop is
\begin{equation}\label{e:refmacro}
a_{ref,m}(z) = a_m(z)\Theta(z)+h_0\Theta(-z).
\end{equation}
The slopes at the intersection of the asymptote $a_m(z)$ and the
homogeneous layer $h(z)\equiv h_0$ at $z=0$ defines the contact
angle $\theta_m(R) = \theta(R,V_{ex}\to\infty)$. $R=\infty$
corresponds to a planar substrate for which the interface profile
diverges \emph{linearly} in the limit $z\to\infty$:
\begin{equation}\label{e:asympplanar}
a_{m,\infty}(z)-R = l_0+z\tan\theta_{\infty}
\end{equation}
with the macroscopic contact angle $\theta_{\infty} =
\theta_m(R\to\infty)$ on the planar substrate. $\theta_{\infty}$ obeys Young's law
$\cos\theta_{\infty} = (\sigma_{wg}-\sigma_{wl})/\sigma$ where
$\sigma_{wg}$ and $\sigma_{wl}$ are the wall-gas and wall-liquid
surface tensions, respectively; $\sigma_{wg}-\sigma_{wl} =
\omega_c(l_0;R=\infty)$ is determined by the effective interface
potential of the corresponding planar substrate (see, c.f.,
Subsec.~\ref{ss:eip}). On the cylindrical surface the contact
angle $\theta_m(R)$ does not follow from similar thermodynamic
considerations but follows from the numerical analysis of the full
profile $h(z)$ for large $V_{ex}$ (see, c.f., Sec.~\ref{s:dsp} and
Fig.~\ref{f:theta_macro}).

The integration constants $C$ and $D$ in Eq.~(\ref{e:asympmacroCandD})
can be determined from the conditions $a_m(z=0) = R+l_0$ and
$a'(z=0) = \tan\theta_m$ so that
\begin{equation}\label{e:asympmacro}
a_m(z) = R\cos\theta_m\,\cosh\left(\frac{z}{R\cos\theta_m}+
\mbox{arccosh}\frac{1}{\cos\theta_m}\right).
\end{equation}
The series expansion of this expression in terms of small $z/R$ is
\begin{equation}\label{e:asympmacro_exp}
a_m(z/R\ll1) = R+l_0+z\,\tan\theta_m(R) + {\mathcal O}(z^2).
\end{equation}
In the limit $R\to\infty$ the region where the higher order terms
are relevant is shifted towards $z=\infty$ such that, with
$\theta_m(R\to\infty) = \theta_{\infty}$, one recovers the linearly
diverging asymptote $a_{m,\infty}(z)$ (Eq.~(\ref{e:asympplanar})). 

\subsection{Effective interface potential}
\label{ss:eip}

For the same liquid layer thickness $l$ the effective interface
potential $\omega_c(l;R)$ of a cylinder differs from that 
of a planar substrate $\omega_p(l)$. The full expression
$\omega_c(l;R)$ is presented in Ref.~\cite{bieker} as obtained from
density functional theory and within a so-called sharp-kink approximation for the
solid-liquid and the liquid-vapor
interface profiles. For reasons of simplicity, here we use the leading order
of a series expansion of $\omega_c(l;R)$ in terms of $d_w/R$ where
$d_w$ is the radial 
extension of the volume excluded for the fluid particles due to the
repulsive part of the substrate potential:
\begin{eqnarray}\label{e:omegac}
\omega_c(l;R) & = &
\frac{3\pi}{2}\,a\,\frac{R}{h^3}\,_2F_1\left(\frac{5}{2},
  \frac{3}{2};2;\left(\frac{R}{h}\right)^2\right) 
+ 8b\,\frac{R}{h^4}\,_2F_1\left(3,2;2;\left(\frac{R}{h}\right)^2\right)
\nonumber\\
& & + \frac{315\pi}{32}\,c\,\frac{R}{h^9}\,_2F_1\left(\frac{11}{2},
  \frac{9}{2};2;\left(\frac{R}{h}\right)^2\right) 
+ {\mathcal O}\left(\frac{d_w}{R}\right)
\end{eqnarray}
with $h=l+R$ and $_2F_1$ hypergeometric functions. In the limit
$l/R\to0$ one recovers the expression for the effective interface
potential of the corresponding planar substrate:
\begin{equation}\label{e:omegap}
\omega_c(l;R\to\infty) = \omega_c(l\to0,R)=
\omega_p(l) = a\,l^{-2}+b\,l^{-3}+c\,l^{-8}.
\end{equation}
However, the power-law decay of $\omega_c(l\to\infty)$ for a fixed,
finite cylinder radius $R$ is
\begin{equation}\label{e:omegac_lr}
\omega_c(l\to\infty;R) = \frac{3\pi}{2}\,a\,\frac{R}{l^3} + {\mathcal O}(l^{-4}),
\end{equation}
i.e., one power faster than that for the corresponding planar substrate.

At present there exists, to our knowledge, only one study concerned
with the shapes of droplets on cylinders within the range of the
effective interface potential between the cylinder surface and the liquid-vapor
interface~\cite{neimark}. However, in Ref.~\cite{neimark} the
\emph{disjoining pressure} $\Pi_c(l) = -(R/(R+l))\,d\omega_c(l)/dl$ on the
right hand side of the Euler-Lagrange equation~(\ref{e:ele}) as a
whole rather than only the effective interface potential $\omega_c$ is 
replaced by the disjoining pressure of the corresponding planar
substrate $\Pi_p(l) = -d\omega_p(l)/dl$. In view of
Eqs.~(\ref{e:omegap}) and (\ref{e:omegac_lr}), except for the factor
$R/(R+l)$, this corresponds to the short-distance expansion ($l/R\to0$)
of the effective interface potential of the cylinder. 
This replacement of the disjoining pressure by that of the planar
substrate is expected to yield numerically reliable results only for large cylinder
radii and small liquid layer thicknesses $l\ll R$. Therefore in
Sec.~\ref{s:dsp} we test the quality of this
approximation (as well as that of the replacement of $\omega_c$ alone
by $\omega_p$). 

So far, due to the volume constraint, our
considerations apply to nonvolatile liquids. For volatile liquids any 
droplet surrounded by a macroscopic reservoir of vapor phase is
thermodynamically unstable 
against evaporation, leaving behind only the thin equilibrium wetting film.
However, we expect that the actual nonequilibrium state of a condensating or
evaporating liquid observed within a time scale that is small compared
with the typical condensation or evaporation time can be described by
solutions of Eq.~(\ref{e:ele}) with $V_{ex}$ given by its momentary
value. Only the interface configuration for 
$\kappa=0$, i.e., $V_{ex}=\infty$, which interpolates between
a homogeneous wetting layer and an exponentially diverging profile, describes a
bona fide thermodynamically stable state which can be maintained by
imposing appropriate
boundary conditions (see above) at liquid-vapor coexistence for the
bulk fluid. The thermodynamic state, which in a grand canonical
ensemble is defined
by temperature and chemical potential, enters parametrically into the actual
values of the effective interface potential $\omega_c$ and the liquid-vapor
surface tension $\sigma$.

\section{Shapes of droplet surfaces and contact angles}
\label{s:dsp}

We solve the Euler-Lagrange equation~(\ref{e:ele}) numerically for
fixed values of $\kappa$ and for a 
given effective interface potential $\omega_c(l)$; the value of
$\kappa$, in turn, determines the excess liquid volume $V_{ex}$ and
allows us to establish the relation $\kappa(V_{ex})$. As boundary
conditions in the case $\kappa<0$ (leading to droplets of finite size)
we use that $h(z)$ must approach the 
wetting layer thickness $h_0$ for large $z$ and that $h'(z=0)=0$. The
distance $L/2$, at which the system is cut off, is
chosen large enough so that $h(z=L/2)$ and $h'(z=L/2)$ attain their asymptotic
values $h_0$ and $0$, respectively, within prescribed accuracy. The reference
profile $a_{ref}(z)$ is then calculated numerically by solving the differential
equation~(\ref{e:eleasymp}) with $\kappa^*$ determined by
Eqs.~(\ref{e:R0}) and (\ref{e:eleasymp}) and
with $a(z=0)=h(z=0)$ and $a'(z=0)=0$, up to the point of intersection of 
$a(z)$ and $h_0$ which defines the coordinate $z_1$; $a_{ref}(z\geq
z_1)=h_0$. The contact angle $\theta$ is determined from
Eq.~(\ref{e:ca}) and, as a crosscheck, from Eq.~(\ref{e:cacarroll}).

In all numerical calculations presented henceforth we set $a=3\sigma
s^2$, $b=-5\sigma s^3$, and $c=3\sigma s^8$ such that $s$ sets the length
scale for the range of the effective interface potential (typically
$s\approx1$nm). We divide both sides of 
Eq.~(\ref{e:ele}) by $\sigma$ so that $\omega(l)/\sigma$ is dimensionless and
$\kappa/\sigma$ has the dimension of an inverse length. Alternatively,
instead of introducing $s$ as above one can choose $\sqrt{a/\sigma}$
as the basic length scale which describes the decay of the effective
interface potential; for our choice $\sqrt{a/\sigma}\approx1.73s$. The
effective interface 
potentials $\omega_c(l;R)$ (Eq.~(\ref{e:omegac})) and $\omega_p(l)$
(Eq.~(\ref{e:omegap})) for the above choice of coefficients are shown in
Fig.~\ref{f:eip}.

As a first example we solve Eq.~(\ref{e:ele})
with the effective interface potential $\omega_p(l)$ of
the corresponding planar substrate (Eq.~(\ref{e:omegap})) and the
potential coefficients given above. Figure~\ref{f:example1} shows the
profile of the droplet surface on a cylinder with radius $R=100s$ for
$\kappa s/\sigma=-0.1$. This choice of $\kappa$ leads
to a small droplet with $V_{ex} \approx 1.46\times10^4 s^3$ (i.e.,
containing roughly $10^7$ fluid particles) whose liquid-vapor interface lies
entirely within the range of the effective interface 
potential. Therefore the deviation of the profile from the asymptote
$a(z)$ extends up to the apex of the droplet. The model effective
interface potential used here resembles a typical interface potential
leading to first-order wetting on a planar substrate~\cite{sdreview}. The droplet
surface crosses the reference profile and, upon approaching the apex of the
droplet, it reaches the reference profile from below.

Carroll~\cite{carroll2} has shown that, in the absence of the
effective interface potential, the axisymmetric droplet
configuration is only stable for
\begin{equation}\label{e:stability}
2\left(\frac{h(0)}{R}\right)^3\cos\theta -
3\left(\frac{h(0)}{R}\right)^2 + 1 > 0,
\end{equation}
i.e., if the droplet is large compared with the diameter of the
cylinder and if the contact angle is small. When the droplet volume
decreases or the contact angle increases the axisymmetric droplet
becomes metastable against a so-called ``roll up'' towards the
``clamshell'' configuration~\cite{adam}. Applying the stability
criterion Eq.~(\ref{e:stability}) to the interface profile shown in
Fig.~\ref{f:example1}, we find that this barrel-type configuration is
possibly metastable towards forming the 
``clamshell'' shape. A definitive statement about the stability would
require to refine the criterion in Eq.~(\ref{e:stability}) by
incorporating the effect of the effective interface
potential. However, the determination of the non-axisymmetric 
``clamshell'' equilibrium shape requires a much larger numerical
effort and is therefore beyond the scope of the present paper. One can
define a critical value $V_{ex,c}$ such that for 
$V_{ex}>V_{ex,c}$ the axisymmetric droplet is stable. Upon increasing
$R$, $V_{ex,c}$ increases, too; $V_{ex,c}\to\infty$ in the limit
$R\to\infty$. Only for 
$V_{ex}=\infty$, i.e., for macroscopic drops, and for contact angles
smaller than $90^{\circ}$ (as stated in Sec.~\ref{s:theory} here we do
not consider the case $\theta>90^{\circ}$) the
rotationally symmetric interface shape is stable for any value of $R$.

Figure~\ref{f:example2} shows the droplet shape for the same choice of
potential parameters and for the same cylinder with $R=100s$, but for
$\kappa s/\sigma=-0.005$. This choice of $\kappa$ leads to a much bigger
droplet with $V_{ex} \approx 1.67\times10^8 s^3$. The apex of
the droplet is located at such a large distance 
from the cylinder surface that the effective interface potential is
almost negligible. Therefore the application of
Eq.~(\ref{e:stability}) is reliable; it yields that this particular
droplet is indeed stable against ``roll-up''. In the vicinity of the cylinder
surface the absolute deviation of the
interface profile from the asymptote is similar to that in Fig.~\ref{f:example1}.
As compared with the situation shown in Fig.~\ref{f:example1}, the
point where $l(z)$ crosses the reference profile $a(z)-R$ is shifted
to the right and lies near the three-phase contact line at
$z=z_1$. For the model effective interface potential used here and in
Fig.~\ref{f:example1}, in the region around the apex of the droplet
the profile lies below the reference profile. These results are in
accordance with the findings for the planar, homogeneous
substrate~\cite{getta,bauer1} with the same type of interface potential.
If, on the other hand, $\omega(l)$ corresponds to a system undergoing
a continuous wetting transition, i.e., exhibiting a single minimum
without a potential barrier, the profile of the droplet shape
approaches its asymptote from the outside without crossing it (see
Fig.~8 in Ref.~\cite{bauer1}).

Figure~\ref{f:example_eipcomp} displays the effect of the replacement
of the effective interface potential of a cylinder $\omega_c(l;R)$
(Eq.~(\ref{e:omegac})) by that of the corresponding planar substrate
$\omega_p(l)$ (Eq.~(\ref{e:omegap})) for a droplet with
$\kappa s/\sigma=-0.005$ so that $l(0)\approx361s$ and $z_1\approx453s$
on a thin cylinder with $R=20s$. This droplet also 
satisfies the stability criterion in Eq.~(\ref{e:stability}). For reasons of
clarity in this figure we have plotted the difference $\Delta h(z)$ between
corresponding profiles instead of the profiles themselves. The influence of
approximating $\omega_c(l;R)$ by $\omega_p(l)$ turns out to be rather small: the
difference $\Delta h(z)$ is at most of the order of $s$. It is
even smaller in the case of smaller droplets and thicker cylinders:
the quality of approximating $\omega_c(l;R)$ by $\omega_p(l)$
improves if the cylinder is thicker and the droplet is smaller. In
Ref.~\cite{neimark} the disjoining pressure of a cylinder
rather than the effective interface potential is replaced by its planar-substrate
counterpart. This corresponds to replacing the term $(R/h)\,d\omega_c/dl$ 
on the right hand side of Eq.~(\ref{e:ele}) by $d\omega_p/dl$. The
dashed line in Fig.~\ref{f:example_eipcomp} shows the effect of this
approximation on the surface profile.
As expected, the quality of this approximation
is worse than the substitution of the effective interface
potential alone, although the difference $\Delta h$ is still of the
order of $s$.

Figure~\ref{f:theta} shows the apparent contact angles for the
examples presented in Figs.~\ref{f:example1} and \ref{f:example2}, as
well as for the same set of potential parameters but with $R=20s$, $R=200s$, and
$R=500s$ as function of the excess liquid volume $V_{ex}$. Upon
increasing $R$ the curves are 
shifted upwards and to the right. Due to the exponential divergence of
the interface profile of a macroscopic drop (which is more pronounced for
smaller $R$, see the discussion of Eqs.~(\ref{e:asympmacro}) and
(\ref{e:asympmacro_exp}) above and, c.f., Fig.~\ref{f:macro}), the
determination of the apparent contact angles for very large
drops is, in particular for thin cylinders, numerically
difficult. However, the data indicate that, for any value of $R$,
$\theta(V_{ex})$ is a monotonously decreasing function, with a
vanishing slope $d\theta/dV_{ex}=0$ at $V_{ex}=\infty$. The differences
between the contact angles for different $R$ are minimal at
$V_{ex}=\infty$. For any $R$ there is a sizeable increase of the
apparent contact angle upon decreasing droplet size.

Figure~\ref{f:theta_botheip} shows the effect of  
replacing $\omega_p$ by $\omega_c$ on the apparent contact angles for
the system with $R=20s$ (compare Fig.~\ref{f:example_eipcomp}). The
difference between the contact angles calculated by using $\omega_c$ and
$\omega_p$ is significant. It is much smaller for thicker
cylinders whose contact angles are displayed in
Fig.~\ref{f:theta}. However, the qualitative functional form of the
dependence of $\theta(V_{ex})$ is not affected by the replacement of
$\omega_c$ by $\omega_p$.

For macroscopic drops (i.e., $\kappa=0$ or, equivalently,
$V_{ex}=\infty$) the Euler-Lagrange equation is solved with the 
initial value $h(z=L_1)=h_0$ and a small initial slope $h'(z=L_1)$ (e.g.,
$h'(z=L_1) = 10^{-8}$); the initial slope $h'(z=L_1)=0$ would yield the
trivial solution $h=h_0$. In order to find the asymptote we determine
the integration constants $C$ and $D$ 
in Eq.~(\ref{e:asympmacroCandD}) such that $a_m(L_2) = h(L_2)$ and
$a_m'(L_2) = h'(L_2)$ where $z=L_2$ is the coordinate up to
which the differential equation is integrated numerically; the system
size $L_2-L_1$ is chosen 
large enough so that upon further increase of the system size $C$ and
$D$ remain unchanged within prescribed accuracy. The contact angle
$\theta_m$ can be inferred from the value of $C$
(Eq.~(\ref{e:asympmacro})). Finally the coordinate system is shifted
laterally such that the intersection
of $a_m(z)$ and $h_0$ define the position $z=0$ (which
corresponds to $z=-z_1$ for $\kappa\neq0$).
As mentioned before, for macroscopic drops the rotationally symmetric
configuration satisfies 
the stability criterion Eq.~(\ref{e:stability}) for any $R$.

The dependence of the liquid-vapor interface profiles of macroscopic
drops on the cylinder radius $R$ is shown in
Fig.~\ref{f:macro} using $\omega_p(l)$ and with the same set of
interaction potential parameters as in the previous examples. In
accordance with Eq.~(\ref{e:asympmacro}), the 
interface profiles for cylinders of finite thickness diverge
exponentially in the limit $z\to\infty$. In the limit $R\to\infty$ the
region where higher-order corrections to the linear behavior
(Eq.~(\ref{e:asympmacro_exp})) are relevant is shifted towards
$z\to\infty$ such that in the limiting case $R=\infty$
corresponding to the planar substrate the linear divergence of the
reference profile is recovered. Figure~\ref{f:theta_macro}(a) displays the apparent
contact angles $\theta_m$ corresponding to the profiles shown in
Fig.~\ref{f:macro} as function of $R$. Upon increasing the cylinder radius $R$,
$\theta_m$ approaches Young's contact angle $\theta_{\infty}$ for the
planar substrate as $\theta_{\infty}-\theta_m \sim
R^{-1}$. 

In Fig.~\ref{f:theta_macro}(b) we show the apparent contact
angles of \emph{small} droplets on a \emph{planar} substrate, i.e., in the limit
$R\to\infty$ but with $V_{ex}<\infty$. In this case the reference
configuration is a spherical cap whose circular base has a radius $z_1$ (see
Fig.~\ref{f:system}(a)). For large droplets $\cos\theta$ reaches
Young's contact angle $\cos\theta_{\infty}$ according to the
Neumann-Boruvka equation~\cite{rowlinsonwidom,neumannboruvka}
\begin{equation}\label{e:nb}
\cos\theta-\cos\theta_{\infty} = -\frac{\tau_{\infty}}{\sigma z_1}
\end{equation}
which allows one to determine experimentally the line tension
$\tau_{\infty}$ of three-phase contact on a planar substrate by
varying the droplet size. Figure~\ref{f:theta_macro}(b) demonstrates
that this linear relationship between $\cos\theta$ and $z_1^{-1}$ is
valid only for $z_1/s\gtrsim500$, i.e., for $z_1\gtrsim
500$nm. From Fig.~\ref{f:theta_macro}(b) one infers that $\cos\theta$
decreases more rapidly than predicted by Eq.~(\ref{e:nb}). This
behavior can be accounted for by an effective line tension
$\tau_{eff}(z_1)$ which due to the circular bending of the three-phase contact
line is \emph{larger} than the value $\tau_{\infty}$ of the
corresponding \emph{straight} 
three-phase contact line. Similar results have been obtained by
Dobbs~\cite{dobbs}.

$\theta$ and $\theta_m$ depend on the liquid-vapor surface tension
$\sigma$ which in turn also exhibits a behavior
$\sigma(R)-\sigma(\infty)\sim R^{-1}$. In accordance with the
discussion in Subsec.~\ref{ss:func}, Fig.~\ref{f:theta_macro} does not
yet take into account this indirect dependence of $\theta$ and
$\theta_m$ on $R$ via $\sigma(R)$.

\section{Line tension}

As long as the size of a droplet is \emph{finite} and \emph{fixed} it
is impossible to extract from the total free energy unambiguously and
in a strict thermodynamic sense a line tension associated with the
three-phase contact lines at the ends of the droplet because there are
arbitrarily many ways to form the total free energy as a sum of
various terms. However,
\emph{well-defined} line tensions emerge as coefficients in the
\emph{size dependence} of the free energy of droplets upon approaching
macroscopic drops.
To this end we consider the limit of large drops, i.e.,
$V_{ex}s^{-3}\gg1$ and $z_1/s\gg1$ (see
Fig.~\ref{f:system}(a)). Within the interface displacement model the
excess free energy in Eq.~(\ref{e:func}) can be rewritten as
\begin{equation}\label{e:decomp}
F_{ex} = \sigma A_b - 4\pi R\omega_c(l_0)z_1-\sigma A_c + {\mathcal L}
\end{equation}
where
\begin{equation}
A_b = 2\pi\int\limits_{-z_1}^{z_1}dz\,a_{ref}(z)\sqrt{1+(a_{ref}'(z))^2}
\end{equation}
is the surface area of the ``barrel'' part of the reference surface
$a_{ref}(z)$, $-z_1\leq z\leq z_1$, $A_c = 4\pi h_0z_1$ is the surface
area of the cylinder with radius $h_0$ and length $z_1$ (see
Fig.~\ref{f:system}(a)), and
\begin{eqnarray}\label{e:L}
{\mathcal L} & = & 2\pi\int\limits_{-\infty}^{\infty}dz\,\Bigg(
\sigma\left(h(z)\sqrt{1+(h'(z))^2}-a_{ref}(z)\sqrt{1+(a_{ref}'(z))^2}\right)
\nonumber\\ 
& & + R\,\Big(\omega(h(z)-R)-\omega(a_{ref}(z)-R)\Big)\Bigg) +
2\pi R\int\limits_{-z_1}^{z_1}dz\,\omega(a_{ref}(z)-R).
\end{eqnarray}
For $V_{ex}\to\infty$, $F_{ex}$ (Eq.~(\ref{e:decomp})) is dominated by
the term $\sigma A_b$ which scales proportional to the surface area of
the drop and thus represents a two-dimensional contribution. The
leading subdominant terms are $-4\pi R\omega_c(l_0)z_1$ and $-4\pi
h_0\sigma z_1$, which scale with the linear dimension $2z_1$ of the
drop representing one-dimensional contributions. Finally, the last
term in Eq.~(\ref{e:decomp}) remains finite for $V_{ex}$, $A_b$, and
$z_1\to\infty$:
\begin{equation}\label{e:decompL}
{\mathcal L}(z_1) = {\mathcal T} + {\mathcal O}(z_1^{-1})
\end{equation}
and thus represents a zero-dimensional contribution. In
its turn ${\mathcal T}$ depends on the cylinder radius $R$ such that
for large $R$ it scales proportional to $R$ which leads to the
following definition of an excess free energy per unit length,
henceforth called ``line tension'', associated with the two contact
lines formed at the ends of the droplet with total length $4\pi R$:
\begin{equation}\label{e:tau}
\tau = \frac{{\mathcal T}}{4\pi R}.
\end{equation}
We note that ${\mathcal T}$ is only well-defined in the thermodynamic
limit $\lim\limits_{A_b,z_1\to\infty}[F_{ex}-\sigma A_b+4\pi
R\omega_c(l_0)z_1+\sigma A_c]$ so that higher order terms, e.g., $\sim
z_1^{-1}$, omitted on the right hand side of Eq.~(\ref{e:decompL}) drop
out. In the following ${\mathcal T}$ is understood to have been
obtained via this procedure. It can be expressed in terms of the
solution of Eq.~(\ref{e:ele}) for $\kappa=0$ and of
$a_{ref,m}(z)$ (Eqs.~(\ref{e:asympmacroCandD}) and (\ref{e:refmacro})):
\begin{eqnarray}\label{e:Lm}
{\mathcal T} & = & 4\pi\int\limits_{-\infty}^{\infty}dz\,\Bigg(
\sigma\left(h(z)\sqrt{1+(h'(z))^2}-a_{ref,m}(z)\sqrt{1+(a_{ref,m}'(z))^2}\right)
\nonumber\\ 
& & + R\,\Big(\omega(h(z)-R)-\omega(a_{ref,m}(z)-R)\Big)\Bigg) +
4\pi R\int\limits_{0}^{\infty}dz\,\omega(a_{ref,m}(z)-R).
\end{eqnarray}
Consequently, ${\mathcal T}$ is twice the characteristic excess free
energy associated with the structure of a macroscopic drop near one of
its ends without interference from the other end. On the other hand,
${\mathcal T}$, and thus $\tau$, are defined for any value of $R$. The
ratio $\tau$ formed in Eq.~(\ref{e:tau}) has the property that in the
limit $R\to\infty$ it reduces to the line tension $\tau_{\infty}$ of
the straight three-phase contact line on the corresponding planar
substrate (see, e.g., Refs.~\cite{indekeu,getta,bauer1,dobbsindekeu}),
which is an experimentally observable quantity (compare
Fig.~\ref{f:theta_macro}(b)).

At this stage one should note that the above considerations tacitly
assume that another thermodynamic limit concerning the total system
size, such as the volume of the surrounding vapor phase and the length
$L$ of the solid cylinder, has already been carried out in advance:
$F(h_0)$ is proportional to $L$ and has been subtracted
before. Moreover, we have not considered the bulk free energy of the
surrounding vapor phase and the bulk free energy of the liquid in the
drop proportional to $V_{ex}$ because they do not enter the
description of the droplet shape in terms of an interface displacement
model. As a careful analysis of the line tension $\tau_{\infty}$
within density functional theory for a volatile liquid at gas-liquid
coexistence shows, in comparison with this more complete theory the
interface displacement model misses a contribution which is
independent of the shape $l(x)$ and is determined by
$\theta_{\infty}$ and $l_0$ (the first term in the sum in 
Eq.~(2.19) in Ref.~\cite{bauer1}, denoted as $\tilde{\tau}$ in
Eqs.~(4.2), (4.3), and (4.5) in Ref.~\cite{getta}). Since, however, this constant
contribution turns out to be numerically much smaller than those
contributions captured by the interface displacement model (see Fig.~15 in
Ref.~\cite{getta}), we have
refrained from determining it for the present, much more complicated
geometry, assuming that the size ratios of these types of
contributions remain roughly the same for the planar and the
cylindrical substrate.

Figure~\ref{f:tau_macro} shows the dependence of the line tension
$\tau$ on the cylinder radius $R$ using the planar effective interface
potential $\omega_p(l)$. $\tau$ is given by $s\sigma$ times a
numerical factor of the order of 1. It decreases monotonously for decreasing $R$
and attains its maximum value $\tau_{\infty}$ for $R\to\infty$ as
$\tau_{\infty}-\tau(R)\sim R^{-1}$. We note that this decrease of
the line tension upon decreasing the radius of curvature $R$ of the contact line is
opposite to the increase of the effective line tension
$\tau_{eff}(z_1)$ observed for a decreasing radius
of curvature $z_1$ of the circular three-phase contact line on a planar substrate as
can be inferred from Fig.~\ref{f:theta_macro}(b). Thus line tensions of
curved three-phase contact lines can be smaller or larger than
the line tension of the corresponding straight contact lines. 

Whereas $\tau_{\infty}$ is
experimentally accessible by monitoring the apparent contact angle of
sessile droplets on a planar substrate as function of the droplet
size, $\tau(R)$ cannot be determined experimentally by direct
observation. The basic reason for this difference is that the length
$2\pi z_1$ of the three-phase contact line of the sessile drop on the
planar substrate can vary as function of the droplet size so that the
optimal shape of the droplet responds to the associated cost $2\pi
z_1\tau_{\infty}$ of the free energy, whereas the excess free energy
$2\pi R\tau(R)$ for the ends of the droplets is a constant
contribution with respect to the droplet size on the cylinder due to
the fixed value of $R$. This, however, holds only for the
``barrel''-type shape of the drop, for which the length of the
three-phase contact lines is fixed. For ``clamshell''-type droplet
shapes the length of the three-phase contact line does depend on the
droplet size so that in this case the line tension will influence the
droplet shape. Nonetheless there are systems for which $\tau(R)$ can
be experimentally relevant. If the cylinder is not a hard solid rod
but consists of a soft material like, e.g., vesicles or tobacco viruses
which float vertically at the liquid-vapor interface of a solvent, the
positive line tension $\tau(R)$ will strangle the object locally,
depending on its restoring elastic forces. According to
Fig.~\ref{f:tau_macro} this tweaking force $-d(\tau(R)R)/dR$ is weaker
for thin cylinders.

\section{Summary}

\begin{figure} % Figure 1
\begin{center}
\epsfig{file=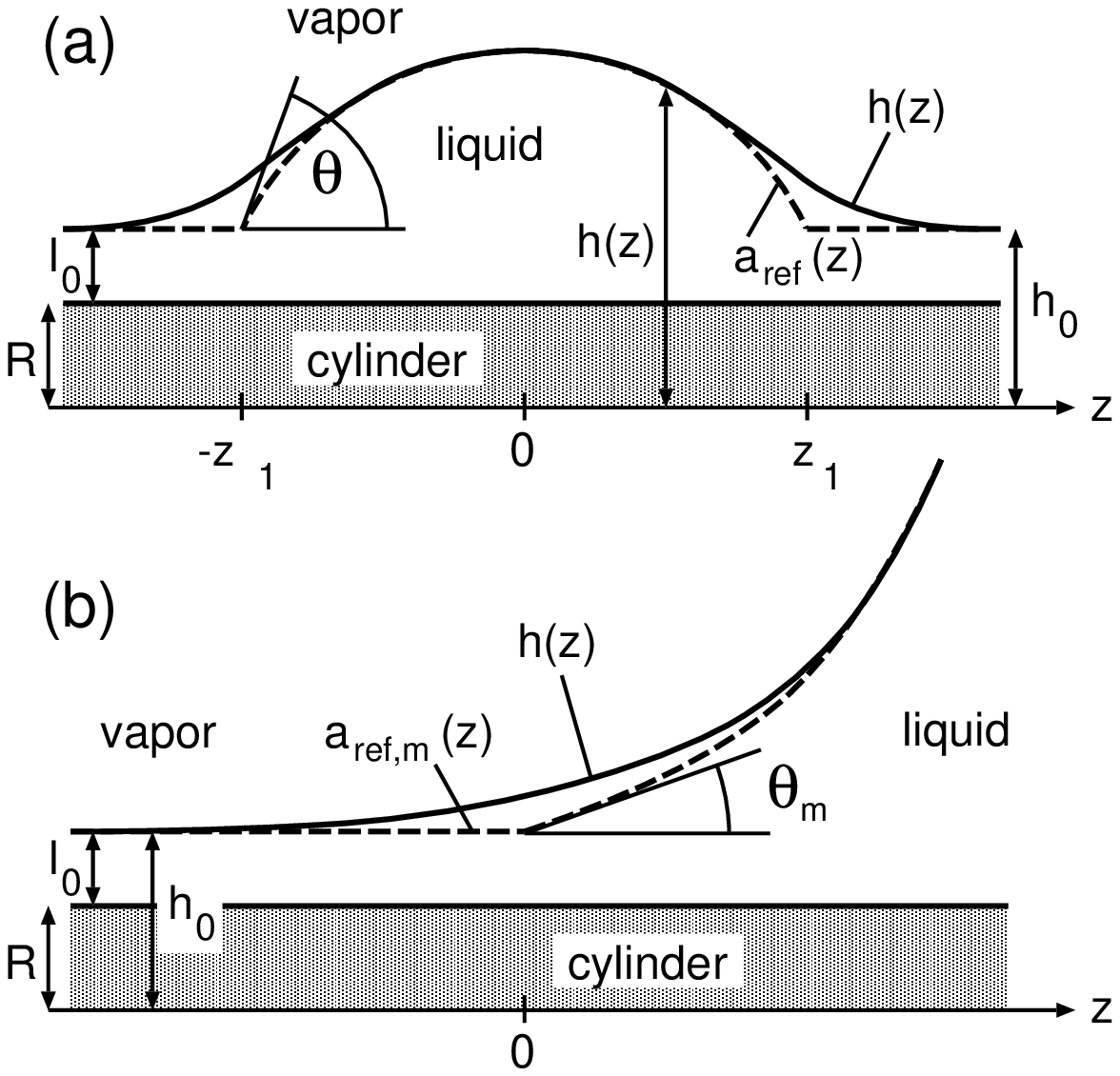, width=9cm}
\end{center}
\caption{\label{f:system}
(a) Schematic longitudinal cross-section through a liquid droplet residing on a
homogeneous cylindrical substrate with radius $R$. The system is
rotationally symmetric around the $z$ axis and symmetric with respect
to a reflection at the plane $z=0$. $h(z)$ (full line) describes the
liquid-vapor interface profile; the thickness of the liquid layer is
$l(z)=h(z)-R$. The droplet shape is determined by the balance of the
Laplace, the capillary, and the disjoining pressure (Eq.~(\ref{e:ele})). Far
from the droplet center, i.e., in the limit 
$|z|\to\infty$, the profile $h(z)$ reduces to a homogeneous layer
$h(|z|\to\infty) = h_0$. $a_{ref}(z)$ (dashed line) describes the
reference surface which consists of a surface $a_{ref}(|z|\leq z_1) =
a(z)$ with constant mean curvature and of the
homogeneous layer $a_{ref}(|z|\geq z_1) = h_0$. $a(z)$ is determined
by $a(z=0)=h(z=0)$ (i.e., it touches the actual surface profile $h(z)$
at the apex) and by the condition that the two principal radii of
curvature $h(0)$ and $R_0$ (which is the radius of curvature of the
planar curve $(z,h(z))$, see Eq.~(\ref{e:R0})) of the actual and the
reference surface at the apex are 
identical. The break in the slope of $a_{ref}(z)$ at
$z=\pm z_1$ defines the apparent contact angle $\theta$. (b) Same as
in (a), but for a macroscopic drop, i.e., infinite excess volume $V_{ex}$.
Choosing the position of one of the contact lines as the origin $z=0$ leads to a
configuration for which the cylinder is in contact with bulk vapor for
$z\to-\infty$ and with bulk liquid for $z\to\infty$. $a_{ref,m}(z\to\infty)$
and $h(z\to\infty)$ diverge exponentially. $\theta_m$ is defined
by the break in the slope of $a_{ref,m}(z)$ at $z=0$.} 
\end{figure}

\begin{figure} % Figure 2
\begin{center}
\epsfig{file=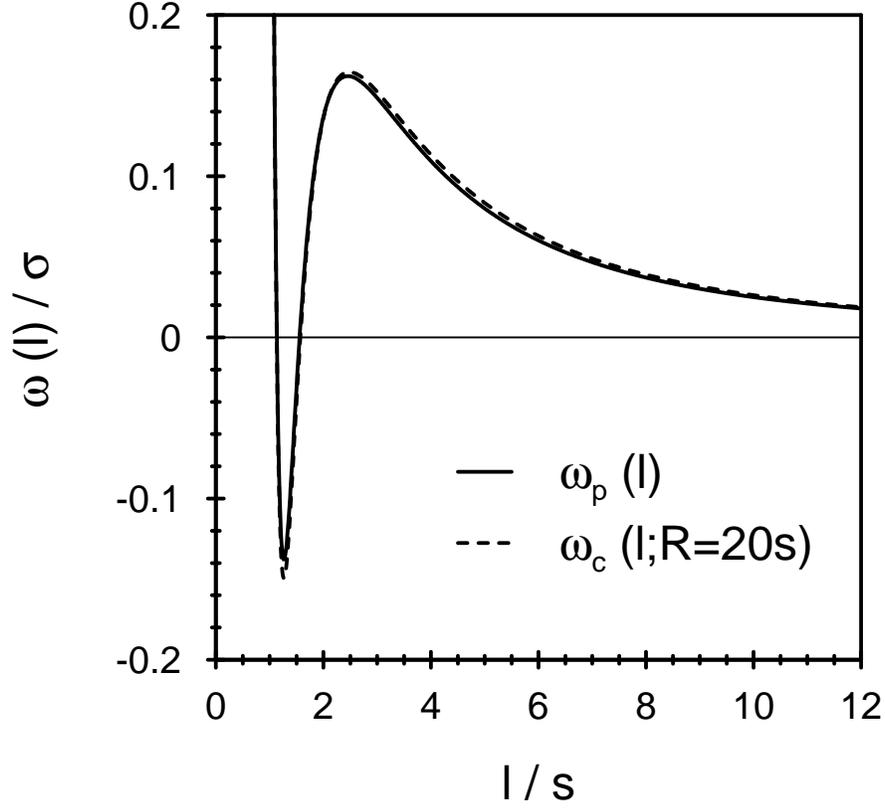, width=12cm, bbllx=50, bblly=340, bburx=525,
  bbury=775}
\end{center}
\caption{\label{f:eip}
Model effective interface potentials $\omega(l)$ in units of the
planar liquid-vapor surface tension $\sigma$ as used in all
numerical calculations. $\omega_c(l;R)$ and $\omega_p(l)$ are given by
Eq.~(\ref{e:omegac}) and (\ref{e:omegap}), respectively, with $a/\sigma = 3s^2$,
$b/\sigma = -5s^3$, and $c/\sigma = 3s^8$ where $s$ sets
the length scale of the system (typically $s\approx1$nm). The full
line denotes the effective interface potential 
$\omega_p(l)$ for the planar substrate, the dashed line
denotes $\omega_c(l;R)$ for a cylinder with $R=20s$.
For this choice of parameters and within the range of values of $l$ shown
here the effective 
interface potentials even of thin cylinders barely differ from
that of a planar substrate. Only for large $l$ the long-range decay of $\omega_p(l)$
and $\omega_c(l)$ differ qualitatively (compare
Eq.~(\ref{e:omegac_lr})). This type of effective interface potential
with a global minimum at $l_0/s\approx1.3$ and a second, local, minimum at
$l=\infty$ leads to a first-order wetting transition of a planar
substrate at a higher wetting transition temperature at which the
first minimum raises up to $\omega=0$. In the case of a continuous
wetting transition $\omega(l)$ would exhibit a single minimum and
approach $\omega=0$ from below in the absence of a potential barrier
in between.}
\end{figure}

\begin{figure} % Figure 3
\begin{center}
\epsfig{file=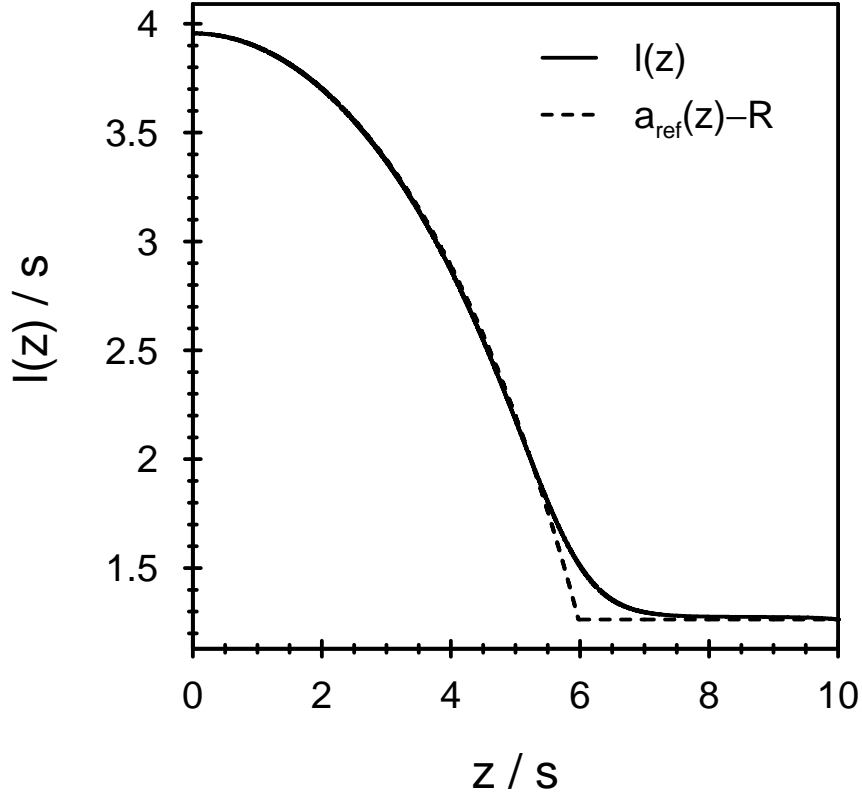, width=12cm, bbllx=50, bblly=340, bburx=525,
  bbury=775}
\end{center}
\caption{\label{f:example1}
Profile $l(z)$ of the droplet surface (full line) and the corresponding
reference profile $a_{ref}(z)-R$ (dashed line) for $R=100s$,
$\kappa s/\sigma=-0.1$ so that $V_{ex}\approx1.46\times10^4 s^3$, and
the effective interface potential $\omega_p(l)$ of 
the corresponding planar substrate as shown in Fig.~\ref{f:eip}. The
droplet is so small that its liquid-vapor interface lies entirely
within the range of $\omega_p(l)$. For $z/s\gtrsim5.25$ the interface
profile $l(z)$ lies above the reference profile 
$a(z)-R$ and, for $z/s\lesssim5.25$ upon approaching the apex of the
droplet, it reaches the reference profile from below. This 
particular droplet is possibly metastable against the ``roll up'' to the
``clamshell'' shape because the stability criterion
Eq.~(\ref{e:stability}) is not satisfied.}
\end{figure}

\begin{figure} % Figure 4
\begin{center}
\epsfig{file=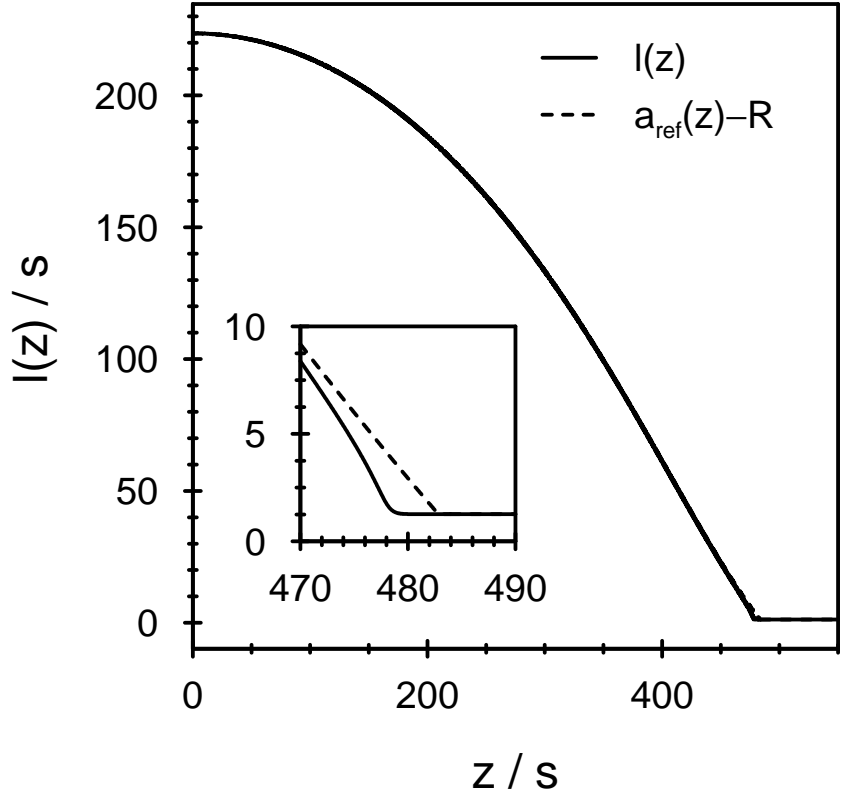, width=12cm, bbllx=50, bblly=340, bburx=525,
  bbury=775}
\end{center}
\caption{\label{f:example2}
Profile $l(z)$ of the droplet surface (full line) and the corresponding
reference profile $a_{ref}(z)-R$ (dashed line) for the same set of
parameters as in Fig.~\ref{f:example1}, but with $\kappa
s/\sigma=-0.005$ so that $V_{ex}\approx1.67\times10^8 s^3$. In
contrast to the situation shown in Fig.~\ref{f:example1}, here the
droplet is so large that it extends up to distances from the
cylinder surface where the effective interface potential is
negligible. The inset magnifies the region around the three-phase
contact line at $z=z_1\approx 483s$. In the region around the apex of
the droplet the profile $l(z)$ lies below the reference profile. Since
the effective interface 
potential is the same as in Fig.~\ref{f:example1} the absolute
deviation of $h(z)$ from $a_{ref}(z)$ is about the same size as in
Fig.~\ref{f:example1} (see the inset). According to
Eq.~(\ref{e:stability}) this droplet is stable against ``roll
up''.}
\end{figure}

\begin{figure} % Figure 5
\begin{center}
\epsfig{file=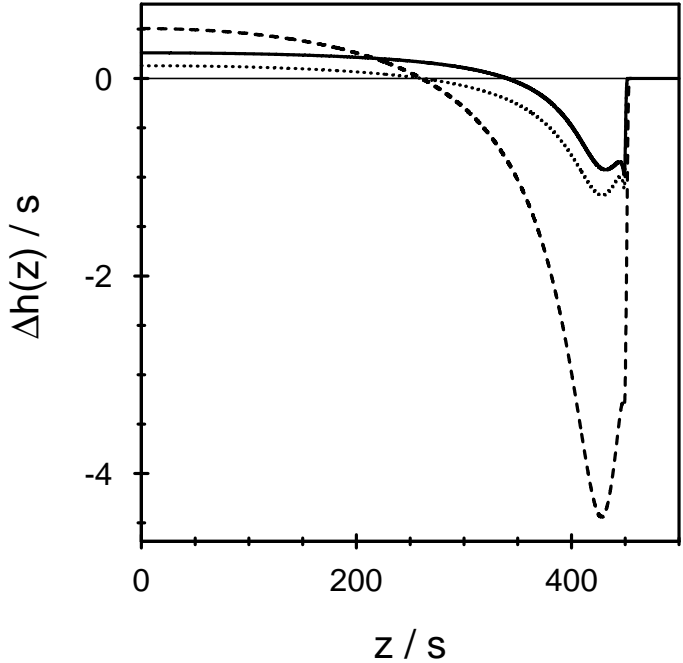, width=10cm, bbllx=50,
  bblly=340, bburx=525, bbury=775}
\end{center}
\caption{\label{f:example_eipcomp}
Full line: difference $\Delta h(z) = h_c(z)-h_p(z)$ between the
two droplet profiles $h_c$ and $h_p$ for $\kappa s/\sigma=-0.005$ (leading to a
droplet with $l(0)\approx361s$ and $z_1\approx453s$) and $R=20\sigma$
which are calculated by using  
$\omega_c(l;R=20s)$ and $\omega_p(l)$, respectively. For
comparison, the dotted line shows the difference between the profile
$h_c$ for $R=20s$ and $\kappa s/\sigma=-0.005$ and the profile $\tilde{h}_p$ for
$R=20s$ but with $\kappa$ chosen such that the excess volumes of liquid
for both profiles $h_p$ and $h_c$ are identical. In order to achieve
this, for the determination of $\tilde{h}_p$ the value of $\kappa s/\sigma$
has to be increased by $1.63\times10^{-6}$. In both cases the maximal
difference is of the order of $s$. Since $\Delta h$
changes sign as function of $z$, it is not possible to find an effective
value $\kappa_{eff}$ such that for a given value of $\kappa$ the
resulting profile calculated with $\omega_p$ is identical with the
profile $h_c$. The dashed line denotes the difference $\Delta h$
between the profile $h_c(z)$ for $\kappa s/\sigma = -0.005$ and $R=20$ and the
profile $\bar{h}_p(z)$ for which, as in Ref.~\protect\cite{neimark}, in
Eq.~(\ref{e:ele}) the entire disjoining 
pressure $(R/(R+l))\,d\omega_c/dl$ (instead of only $d\omega_c/dl$)
is replaced by the disjoining pressure of the planar substrate
$d\omega_p/dl$; here for the
determination of $\bar{h}_p$ the value of $\kappa s/\sigma$ has been
increased by $6.58\times10^{-6}$ in order to have 
identical excess liquid volumes. This approximation
is worse than the substitution of the effective interface
potential alone, although the difference $\Delta h$ is still of the
order of $s$.}
\end{figure}

\begin{figure} % Figure 6
\begin{center}
\epsfig{file=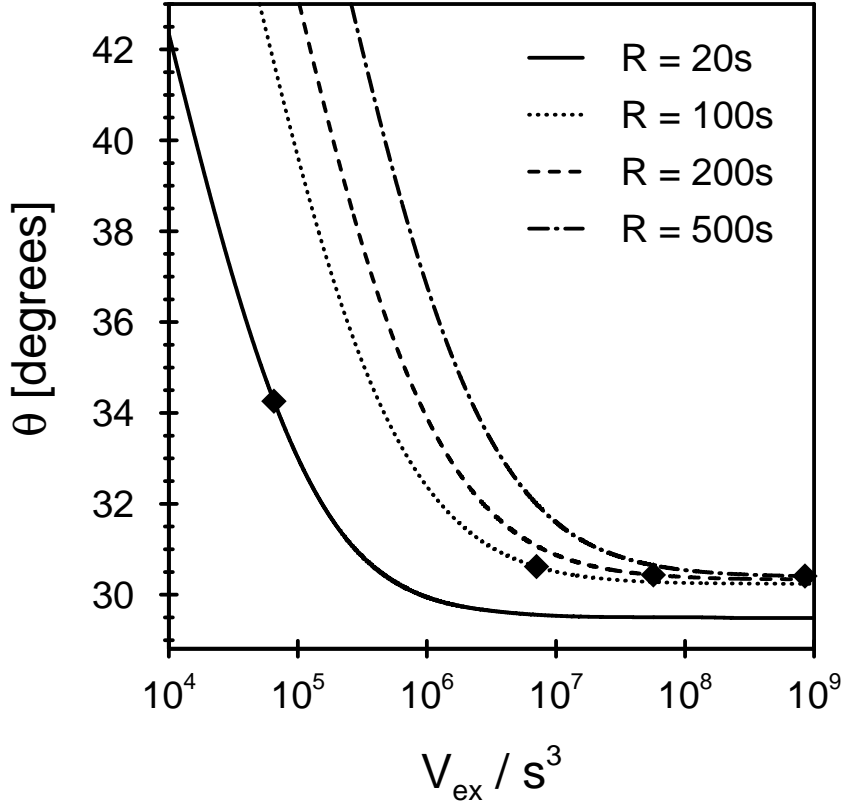, width=12cm, bbllx=50, bblly=340, bburx=525,
  bbury=775}
\end{center}
\caption{\label{f:theta}
Apparent contact angles $\theta$ for the same effective interface
potential $\omega_p(l)$ as used for Figs.~\ref{f:example1} and
\ref{f:example2} and for different cylinder radii $R$ as function of
the liquid excess volume $V_{ex}$. The main effect of increasing
$R$ is a shift of the curves upwards and to the right. The limit
$V_{ex}\to\infty$ corresponds to macroscopic drops or,
equivalently, a cylinder in contact with bulk vapor at one end and
with bulk liquid at the other (compare Fig.~\ref{f:system}(b) and,
c.f., Figs.~\ref{f:macro} and \ref{f:theta_macro}). The differences
between the curves are minimal in the limit
$V_{ex}\to\infty$. $\theta(V_{ex})$ is, for 
any value of $R$, a monotonously decreasing function with vanishing
slope $d\theta/dV_{ex}=0$ in the limit $V_{ex}\to\infty$. The symbols
$\blacklozenge$ 
indicate the critical values $V_{ex,c}$ above which according to the
criterion in Eq.~(\ref{e:stability}) the axisymmetric
droplet shape considered here is stable, but below which the
``clamshell'' configuration is stable.}
\end{figure}

\begin{figure} % Figure 7
\begin{center}
\epsfig{file=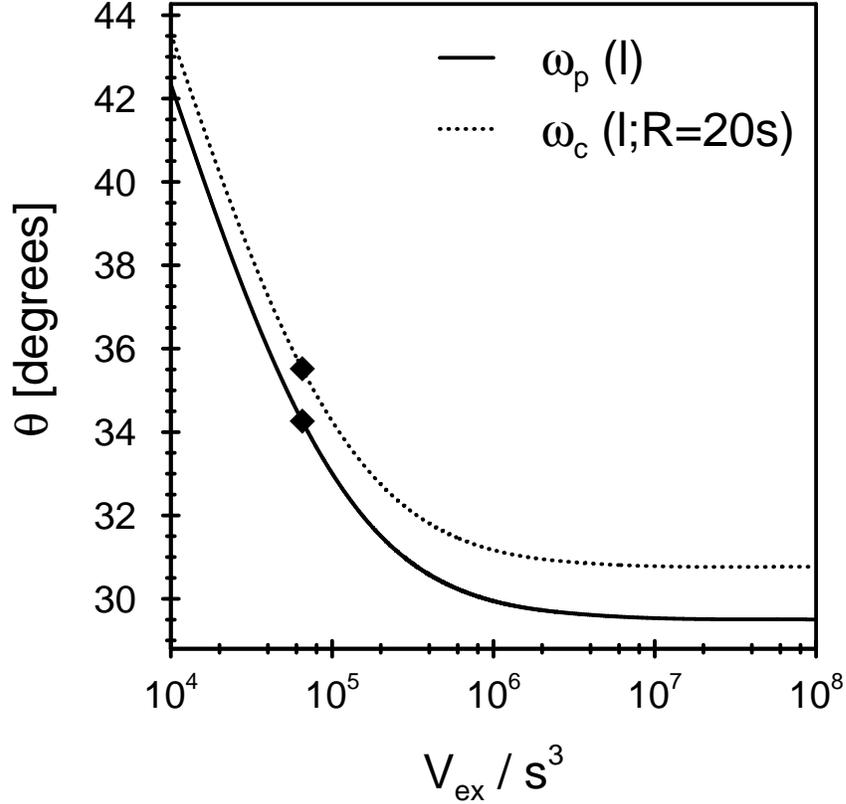, width=12cm, bbllx=50, bblly=340, bburx=525,
  bbury=775}
\end{center}
\caption{\label{f:theta_botheip}
Apparent contact angles $\theta$ calculated by using $\omega_c(l;R)$ (dotted line)
as compared with those calculated based on $\omega_p(l)$ (full line) as
function of the excess liquid volume $V_{ex}$ for a
cylinder with $R=20s$ (i.e., the full line is identical with the full
line in Fig.~\ref{f:theta}). Both curves exhibit the same qualitative
behavior. For thicker cylinders whose contact angles
$\theta(V_{ex})$ are shown in Fig.~\ref{f:theta} the differences between the
curves are much smaller. The symbols $\blacklozenge$
indicate the critical values $V_{ex,c}$ above which the axisymmetric
droplet shape considered here is stable, but below which the
``clamshell'' configuration is stable (compare
Eq.~(\ref{e:stability})). The difference between the two curves is
largest for macroscopicly large drops.}
\end{figure}

\begin{figure} % Figure 8
\begin{center}
\epsfig{file=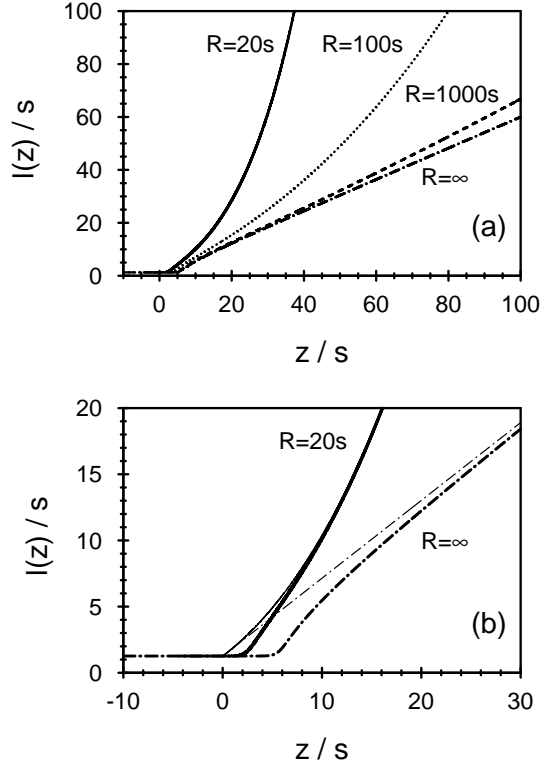, width=7cm, bbllx=75, bblly=100, bburx=525,
  bbury=780}
\end{center}
\caption{\label{f:macro}
(a) Liquid-vapor interface profiles $l(z) = h(z)-R$ of macroscopic
drops ($V_{ex}=\infty$) for $R=20s$ (full line), $R=100s$ (dotted
line), $R=1000s$ (dashed line) and $R=\infty$ (dashed-dotted line),
calculated with $\omega_p(l)$. (b) The profiles $l(z)$ (thick lines) and
their corresponding reference configurations $a_{ref}(z)-R$ (thin lines) for 
$R=20s$ and $R=\infty$ are shown in the magnified region
around the contact line at $z=0$; here we use same graphical
notation as in (a). Each reference configuration consists of the
horizontal line $l(z)=l_0$ for $z<0$ and of the asymptotic branch
$a_m(z)-R$ for $z>0$; for $R<\infty$ this asymptote diverges
exponentially in the limit $z\to\infty$ (see
Eq.~(\ref{e:asympmacro})). $R=\infty$ corresponds to the planar
substrate for which $a_{m,\infty}(z)$ is a \emph{linear}
function. As demonstrated in (a), in the limit $R\to\infty$ the
region where higher-order corrections to the linearly diverging
asymptote become relevant is progressively shifted towards 
$z\to\infty$, such that for $R=\infty$ only
the linear divergence remains. The appertaining contact angles
$\theta_m$ and the line tension $\tau$ as function of $R$ are
shown in, c.f., Figs.~\ref{f:theta_macro} and \ref{f:tau_macro},
respectively. Since for macroscopic drops the contact angles attain a
finite value, for any
radius $R$ the stability criterion in Eq.~(\ref{e:stability}) for the
barrel-type shape is fulfilled.}
\end{figure}

\begin{figure} % Figure 9
\begin{center}
\epsfig{file=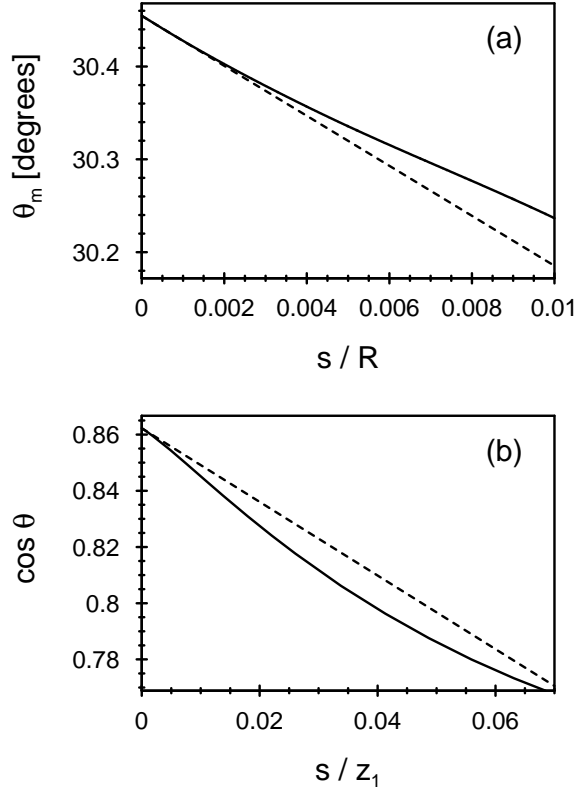, width=8cm, bbllx=30, bblly=135, bburx=525,
  bbury=775}
\end{center}
\caption{\label{f:theta_macro}
(a) Apparent contact angles $\theta_m$ for \emph{macroscopic} drops
($V_{ex}=\infty$) as function of $R$ 
(full line) corresponding to the interface profiles shown in
Fig.~\ref{f:macro}. As indicated by the dashed line, in the limit
$R\to\infty$ the apparent contact angles attain Young's contact angle
$\theta_{\infty}\approx30.5^{\circ}$ on the corresponding
planar substrate as $\theta_{\infty}-\theta_m\sim R^{-1}$. (b)
Apparent contact angle $\cos\theta$ of \emph{small} droplets (full line) on
the \emph{planar} substrate as function of the radius $z_1$ of the
base of the spherical cap acting as the corresponding reference
configuration. For large drops $\cos\theta$ varies according to the
Neumann-Boruvka equation (dashed line) $\cos\theta-\cos\theta_{\infty}
= -\tau_{\infty}/(\sigma z_1)$ which allows one to determine
experimentally the line tension $\tau_{\infty}$ of three-phase contact
on a planar substrate. Here $\tau_{\infty} = 1.31\sigma s$. The
deviation between the full line and the dashed line shows that the
Neumann-Boruvka equation is applicable only for $z_1/s\gtrsim500$.}
\end{figure}

\begin{figure} % Figure 10
\begin{center}
\epsfig{file=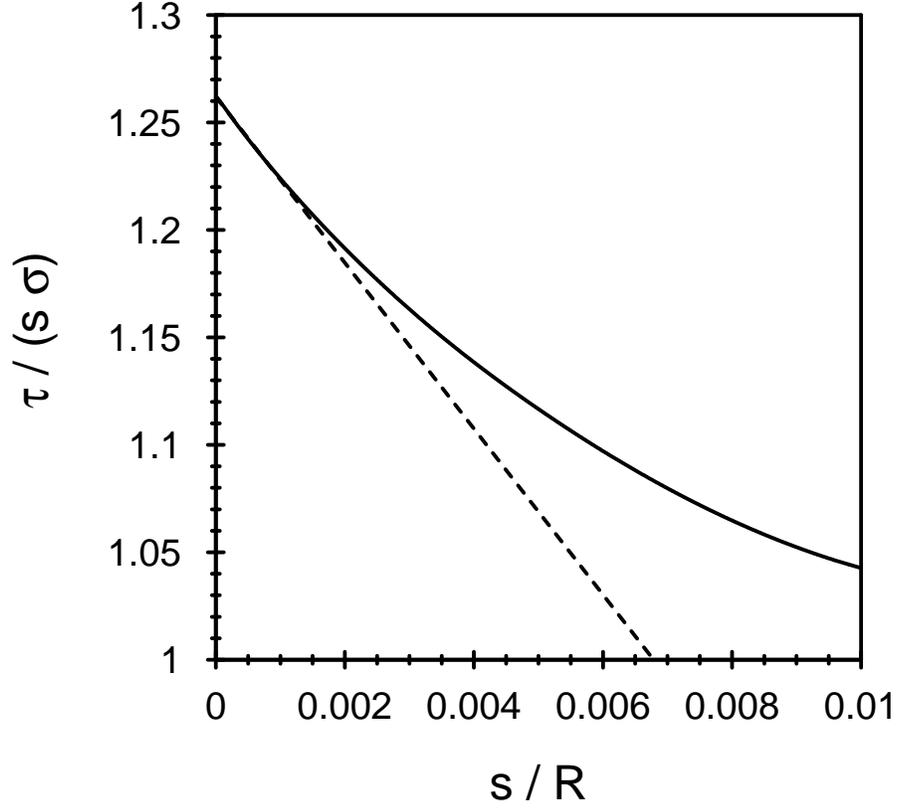, width=12cm, bbllx=50, bblly=340, bburx=525,
  bbury=775}
\end{center}
\caption{\label{f:tau_macro}
Dependence of the line tension $\tau$ (full curve) of macroscopic ``barrel''-type
drops ($\kappa=0$) on cylinders of radius $R$. The corresponding interface 
profiles and apparent contact angles are shown in Figs.~\ref{f:macro}
and \ref{f:theta_macro}, respectively. $\tau$ is calculated by using
the planar effective interface potential
$\omega_p(l)$. As indicated by the dashed line, $\tau$ approaches
$\tau_{\infty}$ for $R\to\infty$ as $\tau_{\infty}-\tau\sim
R^{-1}$. $\tau_{\infty}$ is the line tension of the straight three-phase
contact line on the corresponding planar substrate.}
\end{figure}

Based on an interface displacement model (Eq.~(\ref{e:func})) we have
analyzed the shape and the free energy of ``barrel''-type droplets of
fixed volume $V_{ex}$ covering a cylindrical substrate of radius $R$
(Fig.~\ref{f:system}). For sufficiently small droplets their shape
$h(z) = R+l(z)$ is not only governed by the surface tension $\sigma$ of the
liquid-vapor interface but also by the effective interface potential
$\omega_c(l;R)$ with a generic form as shown in Fig.~\ref{f:eip}.
We have obtained the following main results:

\begin{enumerate}
\item Figures~\ref{f:example1} and \ref{f:example2} show how the
  deviation of the actual droplet shape $h(z)$ from a suitably defined
  reference configuration $a_{ref}(z)$ depends on the droplet
  size. The reference configuration is uniquely defined by the
  requirement to touch the actual shape at the apex and to have a
  constant mean curvature which equals the actual one at the
  apex. $a_{ref}(z)$ allows one to introduce an apparent contact angle
  $\theta$, characterizing the actual shape, which can be expressed in
  terms of the experimentally accessible quantities cylinder radius
  $R$, radii of curvature at the apex, height $l(0)$ of the droplet,
  and thickness $l_0$ of the wetting layer outside the barrel
  (Eq.~(\ref{e:cacarroll})). 
\item The dependence of the effective interface potential
  $\omega_c(l;R)$ on the cylinder radius $R$ influences the shape of
  the droplet on the scale of the range $s$ of $\omega_c(l;R)$
  (Fig.~\ref{f:example_eipcomp}); this dependence has a rather marked
  effect on the apparent contact angle (Fig.~\ref{f:theta_botheip}).
\item The apparent contact angles increase for smaller droplets and
  for thicker cylinders (Fig.~\ref{f:theta}). The contact angles of
  macroscopicly large drops approach Young's contact angle on a planar
  substrate proportional to $1/R$ (Fig.~\ref{f:theta_macro}(a)).
\item In the limiting case of small droplets on a planar substrate the
  circular bend of the three-phase contact line leads to an
  effectively increased value of the corresponding line tension
  (Fig.~\ref{f:theta_macro}(b)). This deviation from the
  Neumann-Boruvka equation becomes relevant if the droplet radius is
  less than roughly $500$nm. This observation is relevant for
  experimental determinations of line tensions via contact angle
  measurements. 
\item For macroscopicly large drops their shape $l(z)$ increases
  linearly on a planar substrate but exponentially on a cylinder
  (Fig.~\ref{f:system}(b)). Figure~\ref{f:macro} illustrates the
  smooth crossover between these types of behavior for increasing
  cylinder radii.
\item For large cylinder radii the line tension associated with the
  ends of macroscopicly large drops approaches the line tension of
  three-phase contact on a planar substrate proportional to $1/R$
  (Fig.~\ref{f:tau_macro}). 
\end{enumerate}

\acknowledgements

This work has been supported by the German Science Foundation within 
the Special Research Initiative \emph{Wetting and Structure Formation
at Interfaces}.

\newpage

\newpage

\end{document}